\documentclass[%
 reprint,
 amsmath,amssymb,
 aps,
 prl,
]{revtex4-1}

\usepackage{graphicx}
\usepackage{dcolumn}
\usepackage{bm}
\usepackage{hyperref}

\usepackage{natbib}
\newcommand{\cqg}{{Class. Quant. Grav.}}
\newcommand{\apjl}{{Astrophys. J. Lett.}}
\newcommand{\apjs}{{Astrophys. J. Suppl. Ser.}}
\newcommand{\jcph}{{J. Comput. Phys.}}
\newcommand{\physrep}{{Phys. Rep.}}
\newcommand{\mnras}{{Mon. Not. R. Astron. Soc.}}
\newcommand{\aap}{Astron. Astrophys.}
\newcommand{\pasp}{Pub. Astron. Soc. Pac.}
\newcommand{\pthphys}{Prog. Theor. Phys.}

\begin{document}


\title{Three-dimensional GRMHD simulations of the remnant accretion disks from neutron star mergers: outflows and r-process nucleosynthesis}

\author{Daniel M. Siegel}
 \altaffiliation[ ]{NASA Einstein Fellow}
\author{Brian D. Metzger}
\affiliation{
 Physics Department and Columbia Astrophysics
  Laboratory, Columbia University, New York, NY 10027, USA
}

\date{\today}

\begin{abstract}
The merger of binary neutron stars, or of a neutron star and a
stellar-mass black hole, can result in the formation of a massive
rotating torus around a spinning black hole.  In addition to providing
collimating media for gamma-ray burst jets, unbound outflows from
these disks are an important source of mass ejection and rapid neutron
capture (r-process) nucleosynthesis.  We present the first
three-dimensional general-relativistic magnetohydrodynamic (GRMHD)
simulations of neutrino-cooled accretion disks in neutron star
mergers, including a realistic equation of state valid at low
densities and temperatures, self-consistent evolution of the electron
fraction, and neutrino cooling through an approximate leakage
scheme. After initial magnetic field amplification by magnetic
winding, we witness the vigorous onset of turbulence driven by the
magneto-rotational instability (MRI). The disk quickly reaches a
balance between heating from MRI-driven turbulence and neutrino
cooling, which regulates the midplane electron fraction to a low
equilibrium value $Y_\text{e} \approx 0.1$. Over the 380 ms duration of the
simulation, we find that a fraction $\approx 20\%$ of the initial
torus mass is unbound in powerful outflows with velocities $v \approx
0.03-0.1\,c$ and electron fractions $Y_\text{e} \approx
0.1-0.25$. Post-processing the outflows through a nuclear reaction
network shows the production of a robust second and third peak
r-process. Though broadly consistent with the results of previous
axisymmetric hydrodynamical simulations,
extrapolation of our results to late times suggests that the total
ejecta mass from GRMHD disks is significantly higher. Our results
provide strong evidence that post-merger disk outflows are an
important site for the r-process.
\end{abstract}

\maketitle


\textit{Introduction.---}Approximately half of the elements heavier than iron are synthesized by the capture of neutrons onto lighter seed nuclei in a dense neutron-rich environment in which the timescale for neutron capture is shorter than the $\beta-$decay timescale 
\cite{Burbidge+57,Cameron57}. This `rapid neutron-capture process' (r-process) occurs along a nuclear path far on the neutron-rich side of the valley of stable isotopes. Despite this realization 70 years ago, the identity of the astrophysical sites giving rise to the r-process remains an enduring mystery \cite{Qian&Wasserburg07,Arnould+07,Thielemann+11}.  

Among the promising r-process sites are the mergers of compact binaries consisting of two neutron stars (NS-NS, BNS; \cite{Symbalisty&Schramm82}) or of a NS and stellar-mass black hole (NS-BH; \cite{Lattimer&Schramm74}). These violent events produce several sources of neutron-rich ejecta, which contribute to their total r-process yields \cite{Rosswog2015,Fernandez2016a}. Historically, most work has focused on matter ejected during the merger process itself, either by tidal forces or due to shock and compression-induced heating at the interface between merging bodies \cite{Rosswog2005,Oechslin2007,Bauswein2013a,Hotokezaka2013b,Kyutoku2015,Kastaun2015a,Radice2016}. While there is broad agreement that a portion of this ``dynamical ejecta" is sufficiently neutron-rich to create the heaviest r-process elements, its quantity is sensitive to the properties of the merging binary and the NS equation of state (EOS).

NS mergers are also accompanied by the formation of a massive accretion disk surrounding the central compact object (e.g., \cite{Ruffert1997,Shibata2006a}). Soon after forming, the neutrino luminosity of the disk is high \cite{Popham1999}, driving a small quantity of mass from the disk surface in a neutrino-driven wind \cite{Surman+08,Metzger+08b,Dessart+09,Perego+14,Richers+15,Martin+15}.  

On longer timescales of hundreds of milliseconds, the disk expands radially due to the outwards transport of angular momentum. One-dimensional models of this spreading evolution using an $\alpha$-prescription for the effective turbulent viscosity \cite{Metzger+08b,Metzger+09a} showed that, as the disk accretion rate drops, the midplane transitions from a neutrino-cooled state to a radiatively-inefficient one \cite{Lee+09,Beloborodov08}. Powerful outflows were predicted following this transition, once heating from turbulent dissipation and nuclear recombination (chiefly alpha-particle formation) are no longer balanced by neutrino cooling.  

These initial models were followed by two-dimensional hydrodynamical simulations of the disk evolution in a pseudo-Newtonian gravitational potential, which also adopted an $\alpha$-viscosity prescription. \cite{Fernandez2013} and \cite{Fernandez2015a} employed an approximate leakage scheme to account for neutrino cooling, and a `light bulb' irradiation model for the neutrino heating, while \cite{Just2015a} used an energy-dependent two-moment closure scheme for the transport of electron neutrinos and antineutrinos. These works found unbound outflows with electron fractions in the range $Y_\text{e} \sim 0.2-0.4$ \cite{Fernandez2013,Just2015a}, sufficient to produce the entire mass range of r-process elements \cite{Just2015a,Wu2016,Lippuner2017a}. The total fraction of the original disk mass unbound in outflows ranged from $\sim\!5\%$ for a non-spinning BH, to $\sim\!30\%$ for high BH spin $\chi_{\rm BH} \simeq 0.95$ \cite{Just2015a,Fernandez2015a}.

Previous simulations of the remnant disk employ a parameterized hydrodynamical viscosity in place of a self-consistent
physical mechanism for angular momentum transport as mediated by the
magneto-rotational instability (MRI) \cite{Balbus&Hawley98}.
\cite{Shibata2007a} performed two-dimensional general-relativistic magnetohydrodynamic (GRMHD) simulations of the disk evolution lasting 60 ms; however, they were not focused on nucleosynthesis and their restriction to 2D precluded a study of the saturated MRI due to the anti-dynamo theorem.  In this Letter, we present the first fully three-dimensional GRMHD simulations of the remnant accretion disk evolution and its outflows over a timescale of $\approx\!400$ ms following the merger.


\begin{table}[tb]
\caption{Initial configuration: BH mass and 
  dimensionless spin, torus mass, inner and outer torus radius, radius
  at maximum density, specific entropy, electron fraction, and maximum
  magnetic-to-fluid pressure ratio.}
\label{tab:BH_torus}
\centering
\begin{tabular}{ccccccccc}
\hline\hline
 $M_\mathrm{BH}$ & $\chi_\mathrm{BH}$ & $M_\mathrm{t0}$&
 $R_{\mathrm{in},0}$ & $R_{\mathrm{out},0}$ & $R_0$ & $s_0$
  &$Y_{\mathrm{e}0}$ & $p_b/p_\mathrm{f}$\\
$[M_\odot]$ &  & $[M_\odot]$& $[M_\mathrm{BH}]$ & $[M_\mathrm{BH}]$ & $[\mathrm{km}]$ & $[k_\mathrm{B}/\mathrm{b}]$ & \\
\hline
$3.00$ & $0.8$ & $0.03$ & $4$  & $24$ & $30$ & 8 & $0.1$ & $ <5\times 10^{-3}$\\
\hline
\end{tabular}
\end{table}

\textit{Numerical setup and initial conditions.---}Simulations are
performed in ideal GRMHD with a fixed background spacetime using the open-source
\texttt{EinsteinToolkit}\footnote{\url{http://einsteintoolkit.org}}
\cite{Loeffler2012} with the GRMHD code
\texttt{GRHydro} \cite{Moesta2014a}. GRMHD is implemented using a
finite-volume scheme with piecewise parabolic reconstruction \cite{Colella1984}, the HLLE Riemann solver
\cite{Harten1983,Einfeldt1988}, and constrained transport \cite{Toth2000}
for maintaining the magnetic field divergence-free. We have implemented a
new framework for the recovery of primitive variables in \texttt{GRHydro} that provides support for any 3-parameter
EOS, as well as a recovery scheme based
on three-dimensional root finding according to \cite{Cerda-Duran2008}, which shows better and faster convergence
properties than two-dimensional schemes; its ability to
recover strongly magnetized regions is important for evolving
low-density, magnetized disk winds.

Thermodynamic properties of matter are described by the Helmholtz EOS
\cite{Timmes1999,Timmes2000}, which includes contributions to the
Helmholtz free energy from nuclei (treated as ideal gas) with Coulomb
corrections, electrons and positrons with an arbitrary degree of
relativity and degeneracy, and photons in local thermodynamic
equilibrium. We consider free neutrons,
protons, and alpha particles, whose abundances are calculated assuming nuclear statistical equilibrium (NSE). We add
dissociation energies to the Helmholtz EOS as in \cite{Fernandez2013} to account for the energy release from alpha-particle formation, as well as the additional terms to the thermodynamic derivatives arising from compositional changes.

\begin{figure}[tb]
\centering
\includegraphics[width=0.49\textwidth]{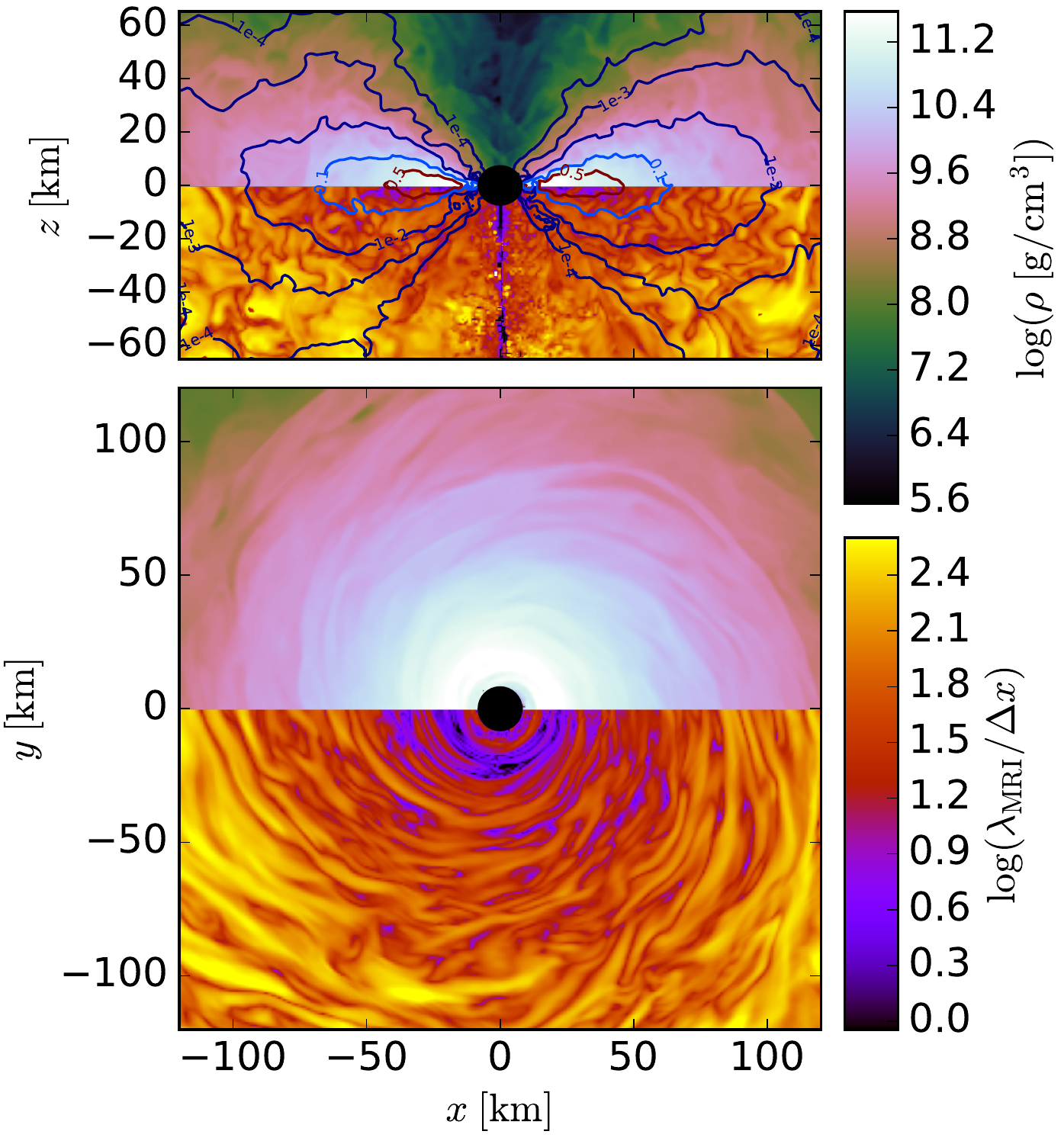}
 \caption{Snapshots of rest-mass density, number of grid points per
   fastest-growing MRI mode, and contours of optical depth to electron
   neutrino number emission $\tau_{\nu_\text{e}}=0.5,0.1,10^{-2},10^{-3},10^{-4}$ at
   $t=20\,\mathrm{ms}$, when the disk has
   settled into a quasi-stationary state (the BH interior is masked).
  }
  \label{fig:snapshots1}
\end{figure}

Neutrino cooling is described by a leakage scheme newly
implemented into \texttt{GRHydro}. Leakage schemes
are widely used in both core-collapse supernovae and
compact-binary merger simulations (e.g., \cite{vanRiper1981,Ruffert1996b,Rosswog2003b,Sekiguchi2011,Ott2013,Perego2016}). 
Our implementation follows closely \cite{Radice2016}, which is based on
\cite{Galeazzi2013} and employs the formalism by
\cite{Ruffert1996b}. We calculate optical depths following the
procedure by \cite{Neilsen2014}, which is well suited for the
aspherical geometry of an accretion disk. We neglect neutrino
absorption, which is expected to appreciably change the outflow dynamics only for significantly more massive accretion
disks (\cite{Fernandez2013}; see also Fig.~\ref{fig:snapshots1}).

Initial data consists of an equilibrium torus of constant specific
angular momentum and specific entropy around
a Kerr BH \cite{Stergioulas2011c,Friedman2013}
(Tab.~\ref{tab:BH_torus}). We compute a torus solution in
horizon-penetrating Kerr-Schild coordinates, which we use in our
simulation. The BH mass and spin correspond to a typical NS merger
remnant. BHs promptly formed in BNS
mergers show spins $\chi_\mathrm{BH}\approx 0.8$
\cite{Kiuchi2009,Rezzolla2010,Bernuzzi2014}, and are
unlikely to be significantly larger \cite{Kastaun2013}, whereas BHs
formed by delayed collapse typically
show spins $\chi_\mathrm{BH}\lesssim 0.7$
\cite{Sekiguchi2016}. Furthermore, $\chi_\mathrm{BH}\sim 0.8$ also
represents a typical BH spin for BH--NS mergers required to tidally disrupt the
NS and form a sufficiently massive torus \cite{Foucart2012}. The
initial torus mass also corresponds to typical NS merger scenarios (e.g.,
\cite{Hotokezaka2011,Foucart2017}). We set up a weak initial magnetic
seed field inside the torus with vector potential components
$A^r=A^\theta = 0$ and $A^\phi =
A_b\, \mathrm{max}\{p-p_\mathrm{cut},0\}$. Here, $p$ denotes the fluid
pressure, $p_\mathrm{cut}=1.3\times 10^{-2}p_\mathrm{max}$, where
$p_\mathrm{max}$ is the pressure at maximum density in the torus, and
$A_b$ sets the initial field strength, which we adjust such that the
maximum magnetic-to-fluid pressure ratio in the torus is $<5\times
10^{-3}$; this results in a maximum magnetic field strength of
$\approx\!3.3\times 10^{14}\,\mathrm{G}$. The torus is initially embedded in a
uniform, tenuous atmosphere with $\rho\approx 37\,\text{g}\,\text{cm}^{-3}$, $T =
10^5\,\text{K}$, and $Y_\mathrm{e}=1$.  This density and
temperature are sufficiently low to neither impact the dynamics nor
the composition of the disk outflows.

Simulations are performed in full 3D without employing symmetries. The
grid setup consists of a Cartesian grid hierarchy of 8 refinement levels,
extending from the center of the BH to $1.53\times
10^9\,\mathrm{cm}$ in every coordinate direction. The finest refinement level
corresponds to a spatial domain with a resolution of $856\,\mathrm{m}$ and
a diameter of $240\,\mathrm{km}$, which entirely contains the initial
accretion torus.


\begin{figure}[tb]
\centering
\includegraphics[width=0.49\textwidth]{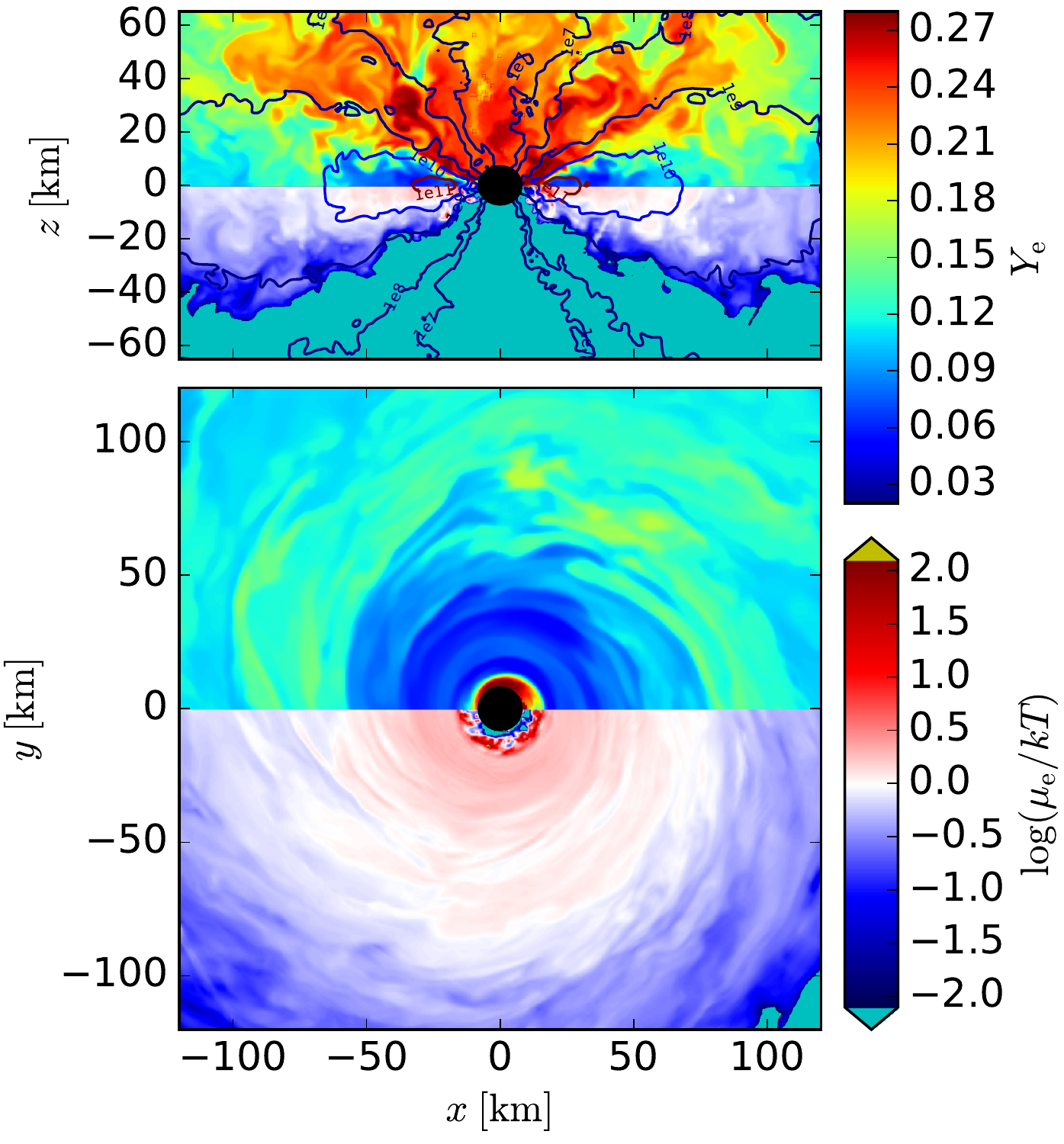}
\caption{Snapshots of electron fraction, normalized electron chemical potential, and
   contours of rest-mass density $\rho=[10^{7},10^{8},10^{9},10^{10},10^{11}]\,\text{g}\,\text{cm}^{-3}$ at
   $t=43\,\mathrm{ms}$, when the disk
   has fully self-regulated itself to mild electron degeneracy (the BH interior is masked).}   
  \label{fig:snapshots2}
\end{figure}

\begin{figure}[tb]
\centering
\includegraphics[width=0.49\textwidth]{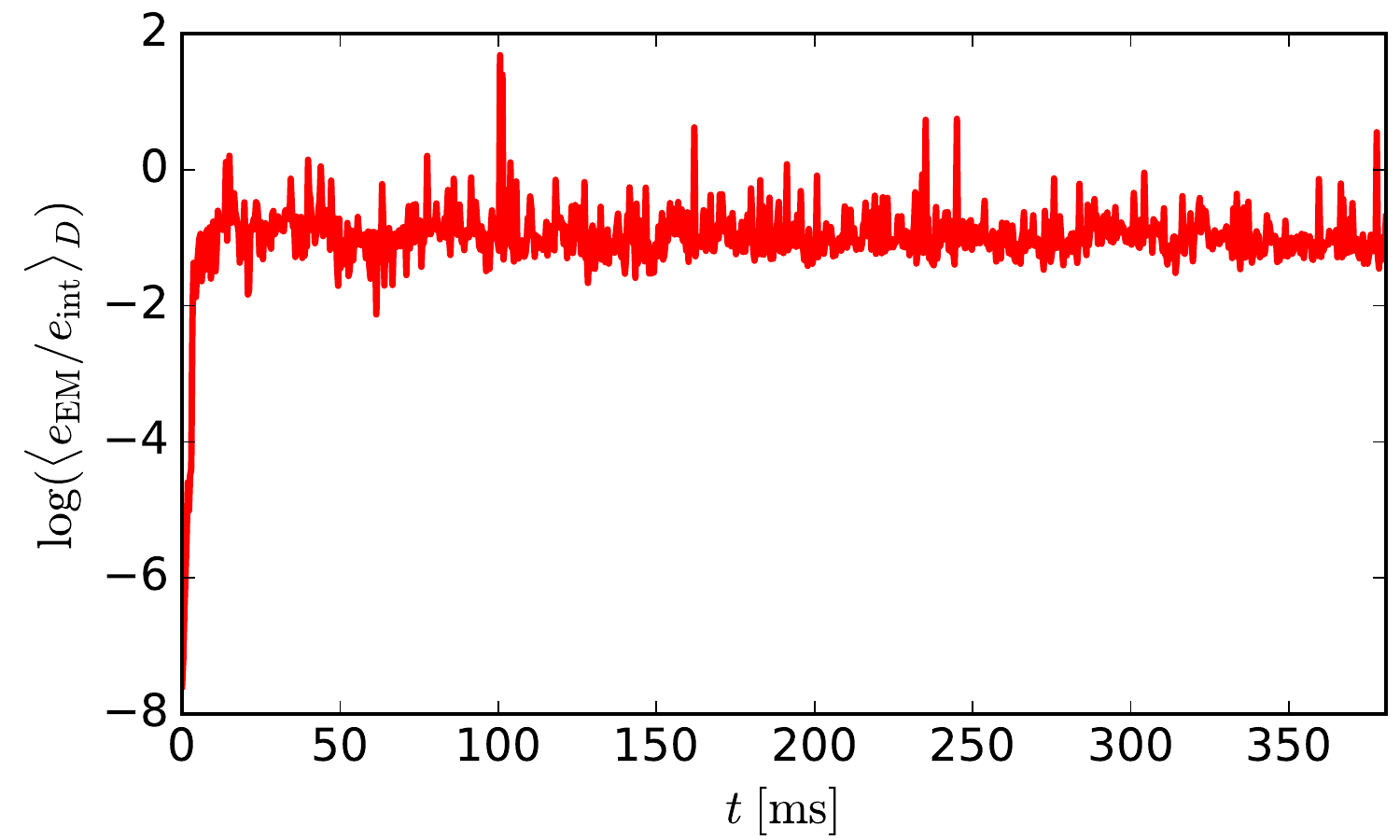}
\caption{Evolution of the density-averaged ratio of
  electromagnetic energy to internal energy in the disk.}
  \label{fig:energy_ratio}
\end{figure}

\begin{figure*}[tb]
\centering
\includegraphics[width=0.24\textwidth]{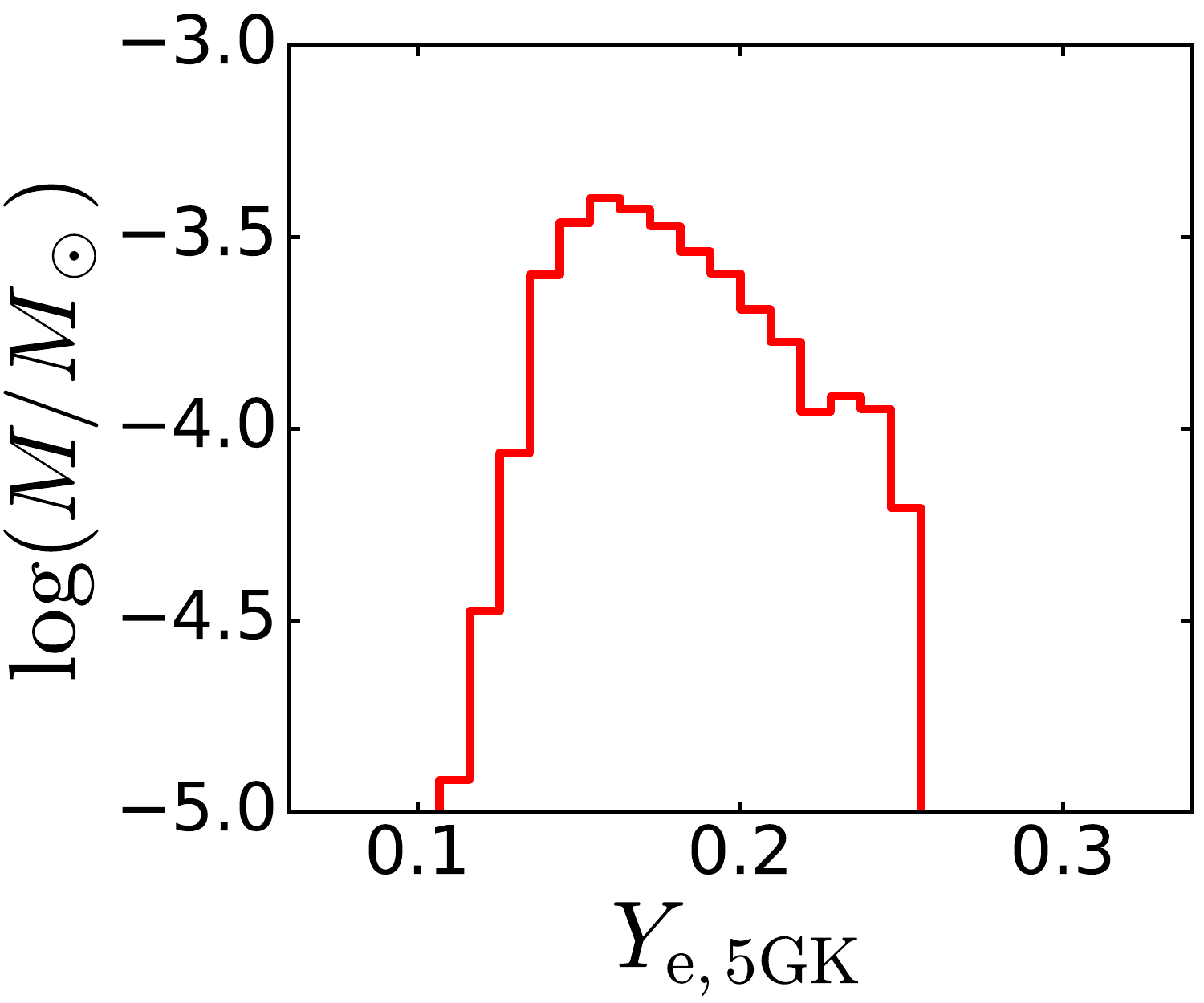}
\includegraphics[width=0.24\textwidth]{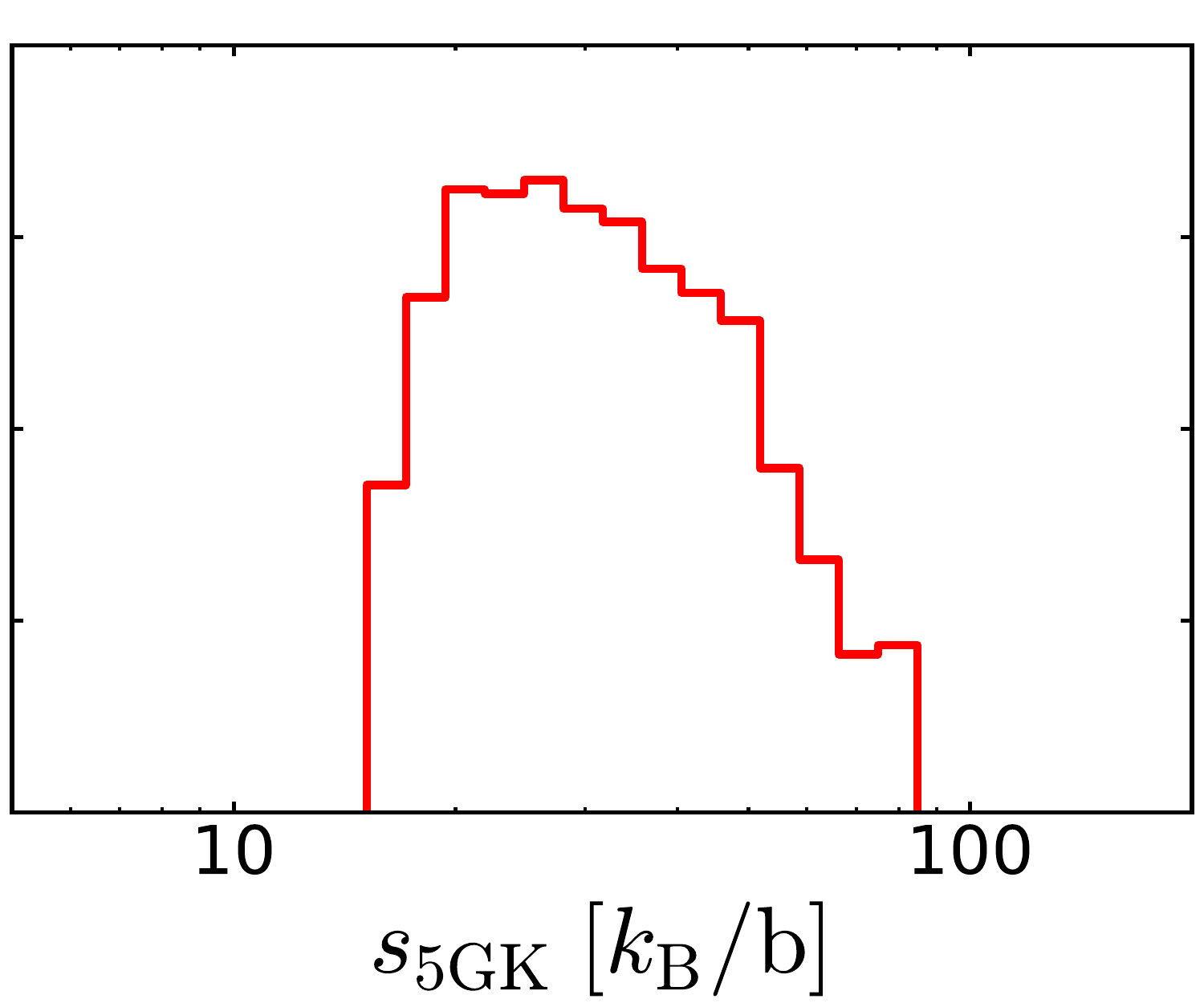}
\includegraphics[width=0.24\textwidth]{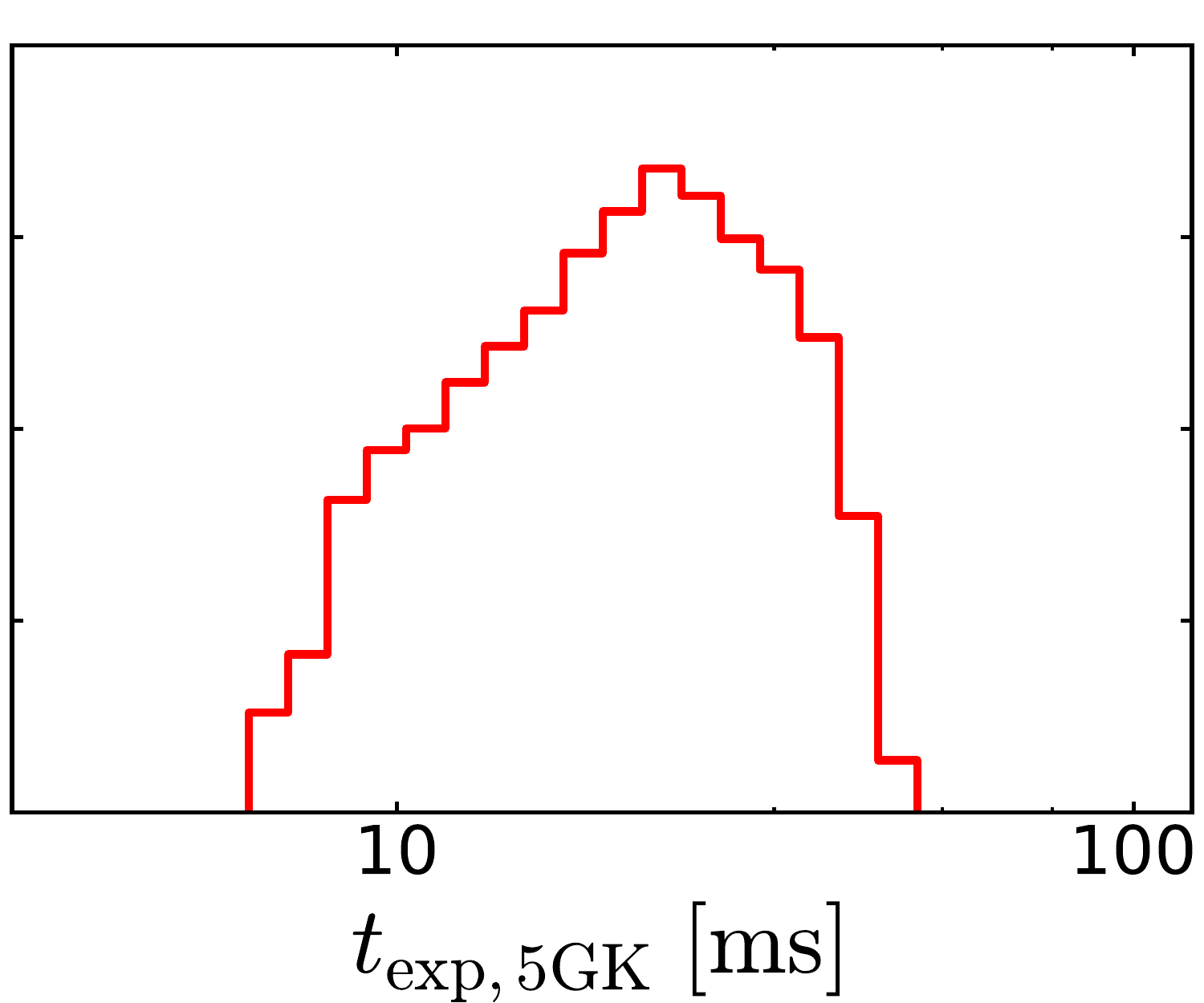}
\includegraphics[width=0.24\textwidth]{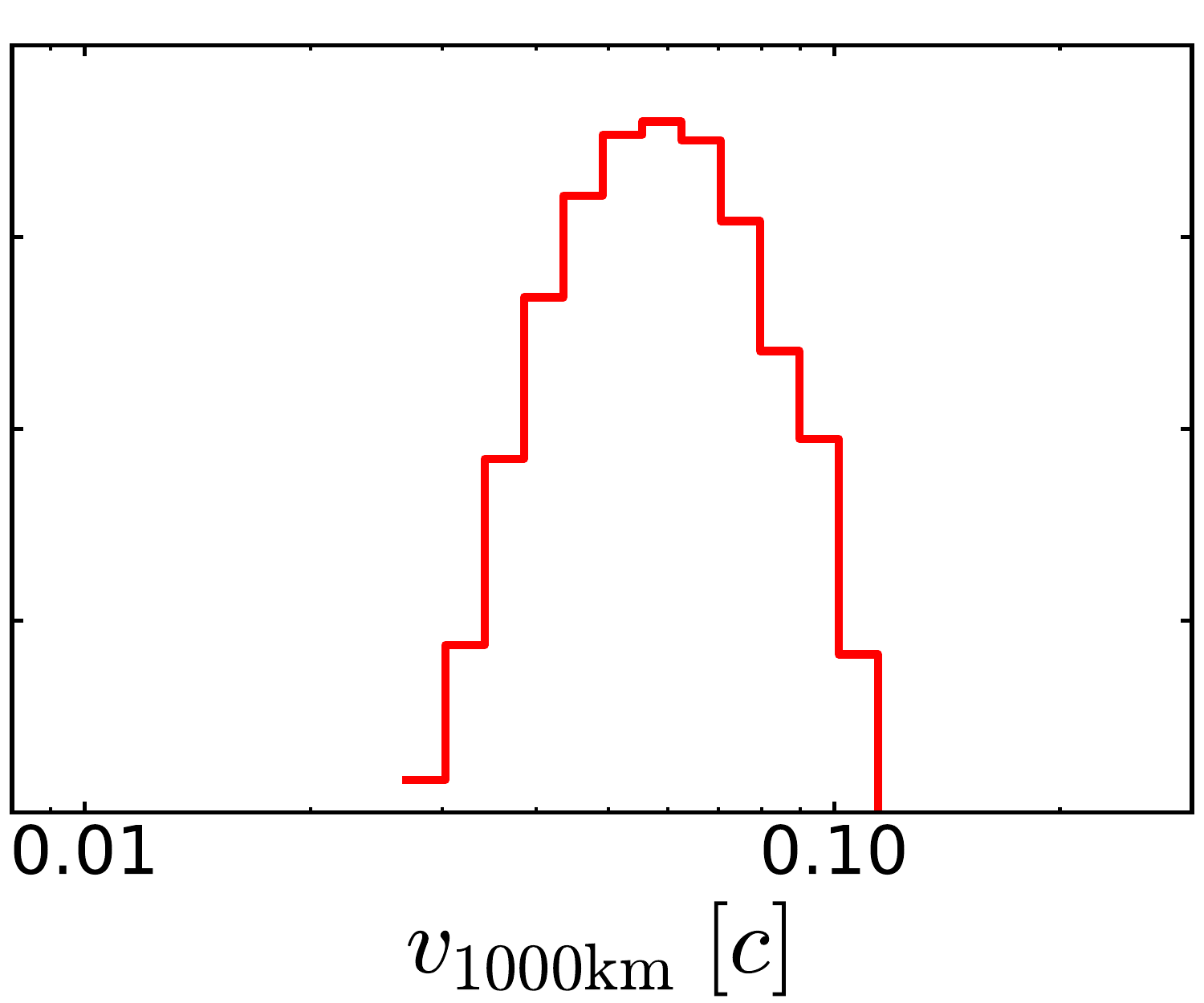}
 \caption{Mass distributions of the unbound disk outflow as measured by tracer particles in terms of electron fraction, specific entropy, expansion timescale (all at $t=t_\mathrm{5GK}$), and outflow velocity at $r=10^3\,\mathrm{km}$.}
 \label{fig:histograms}
\end{figure*}

\begin{figure}[tb]
\centering
\includegraphics[width=0.49\textwidth]{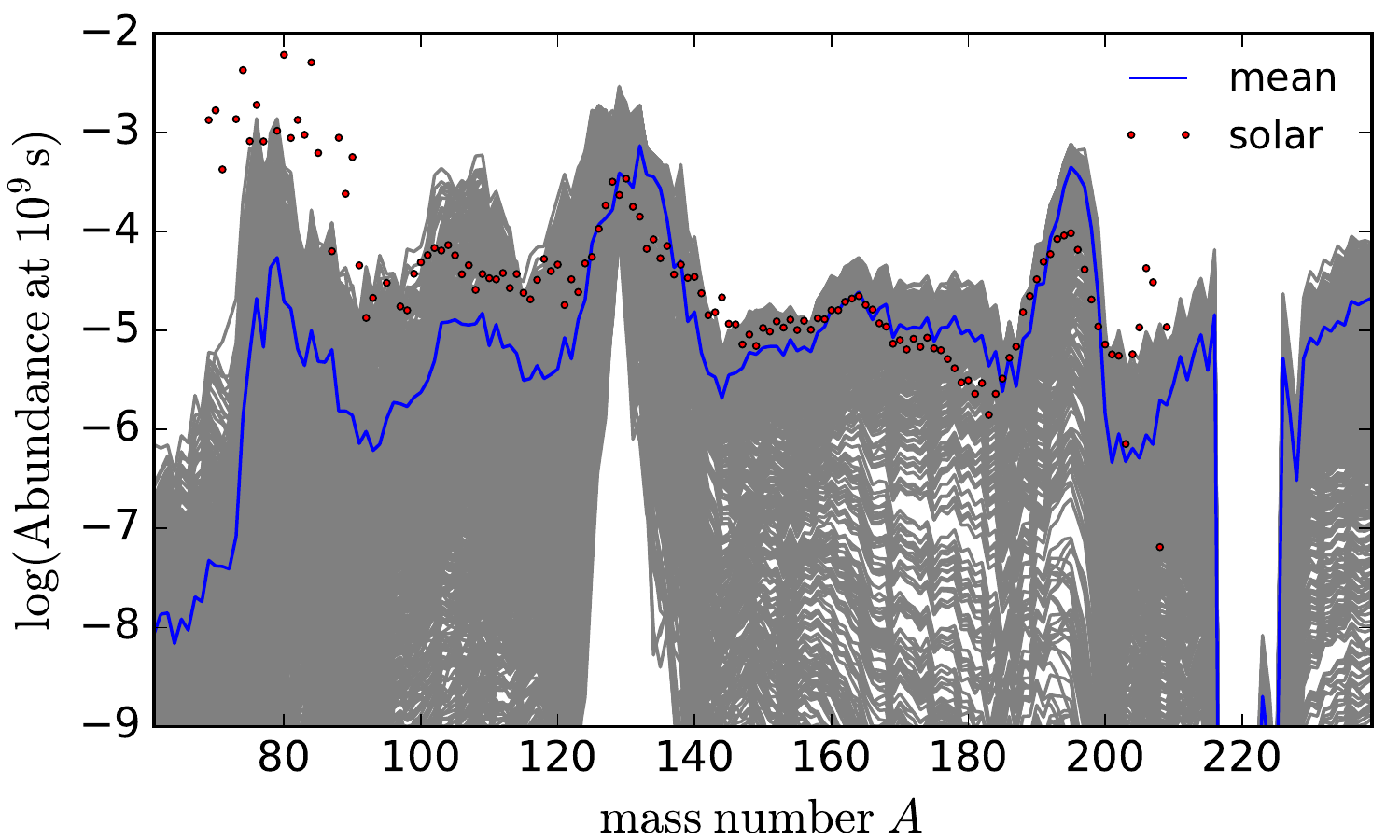}
\includegraphics[width=0.49\textwidth]{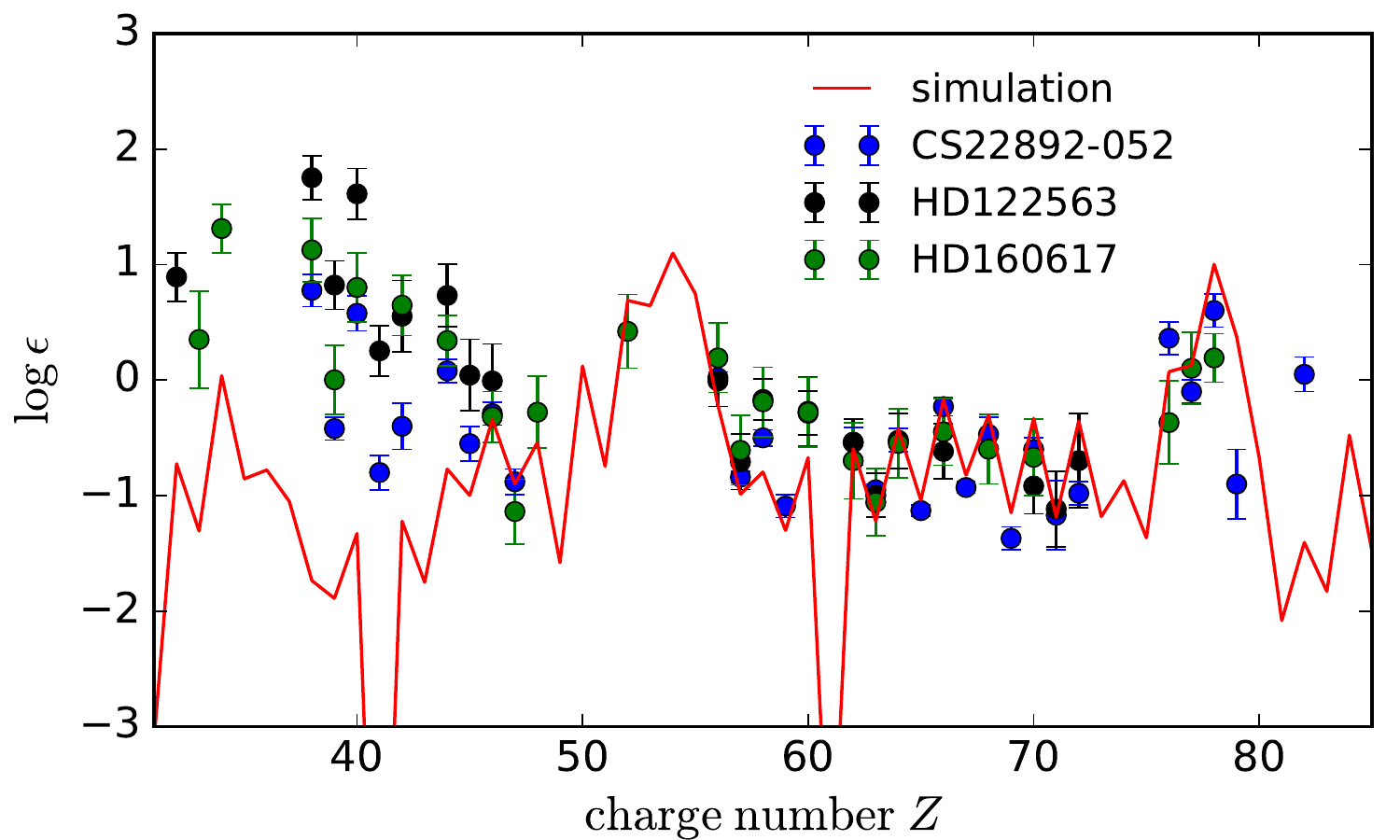}
 \caption{Top: Final elemental abundances from individual unbound
   tracer particles (gray lines) and their mean (blue line),
   compared to observed solar system abundances \cite{Arnould2007},
   scaled to match the mean at $A=130$. Bottom: Comparison
   of the mean abundances to observed abundances in metal-poor halo
   stars \cite{Sneden2003,Roederer2012a,Roederer2012b} in terms of
   $\log\epsilon = \log Y_Z/Y_1 + 12$, scaled such that $\sum (\log Y_Z/Y_{Z,\mathrm{CS22892-052}})^2$ is minimized between $55\le Z \le 75$.}
 \label{fig:nucleosynthesis}
\end{figure}

\begin{table*}[tb]
\caption{Initial torus mass after early relaxation and characteristics
  of the disk outflows as
  measured by tracer particles: mean electron fraction, specific
  entropy, expansion timescales, all at $t=t_\mathrm{5GK}$
  and subdivided into equatorial ($30^\circ < \theta < 150^\circ$) and
  polar ($\theta \leq 30^\circ$ and $\theta \geq 150^\circ$) outflow
  (the polar angle $\theta$ being measured at the end of the
  simulation), as well as total integrated outflow mass
  (polar/equatorial and total). Corresponding
  values by \cite{Fernandez2015a} (F15) and \cite{Just2015a} (J15) are also listed.}
\label{tab:results}
\centering
\begin{tabular}{cccccccccccccc}
\hline\hline
 simulation & outflow type & & \multicolumn{3}{c}{equatorial outflow} & \multicolumn{3}{c}{polar outflow} & \multicolumn{3}{c}{total outflow} & \\
 & & $M_\mathrm{t,in}$ & $\bar{Y}_{\mathrm{e}}$ & $\bar{s}$ & $\bar{t}_\mathrm{exp}$ & $\bar{Y}_{\mathrm{e}}$ & $\bar{s}$ & $\bar{t}_\mathrm{exp}$ & $\bar{Y}_{\mathrm{e}}$ & $\bar{s}$ & $\bar{t}_\mathrm{exp}$ & $M_\mathrm{pol}$ & $M_\mathrm{out}$ \\
 & & $[10^{-2}M_\odot]$ & & 
 $[k_\mathrm{B}/\mathrm{b}]$ & $[\mathrm{ms}]$ & &  $[k_\mathrm{B}/\mathrm{b}]$ & $[\mathrm{ms}]$ & &
 $[k_\mathrm{B}/\mathrm{b}]$ & $[\mathrm{ms}]$ & 
 $[M_\mathrm{eq}]$ & $[M_\mathrm{t,in}]$ \\
\hline
this work & unbound & 2.02 & 0.18 & 31 & 24 & 0.19 & 39 & 18 & 0.18 & 32 & 23 & 0.22 & 0.16 \\
this work & total & 2.02 & 0.17 & 28 & 26 & 0.19 & 43 & 18 & 0.17 & 30 & 25 & 0.15 & 0.23\\
\hline
F15 t-a80 & total & 3.00 & 0.22 & 21 & 35 & 0.31 & 38 & 9.4 & - & - & - & 0.01 & 0.17  \\
J15 M3A8m03a2 & total & 3.00 & - & - & - & - & - & - & 0.27 & 30 & - & - & 0.23 \\
J15 M3A8m03a5 & total & 3.00 & - & - & - & - & - & - & 0.25 & 33 & - & - & 0.24 \\
\hline
\end{tabular}
\end{table*}

\textit{Results.---}The initial torus is evolved from $t=0$ to
$t=381\,\text{ms}$, after which an appreciable fraction of the initial
torus mass has been unbound in powerful outflows. After an initial
transient phase of about $20\,\text{ms}$ due to the onset of
turbulence created by the MRI, the disk settles into a
quasi-stationary state for the rest of the simulation. During
this early relaxation, $\approx\!33\%$ of the initial torus mass
is either accreted onto the BH or ejected via outflows, leaving an
effective initial torus of $\approx\!0.02\,M_\odot$
(Tab.~\ref{tab:results}). We exclude matter ejected or accreted during this
phase from all further analysis.

Figure~\ref{fig:snapshots1} shows snapshots
of several quantities at the beginning of the
quasi-stationary evolution phase. Until the end of the
simulation, the disk and outflows remain qualitatively similar as
depicted here. In particular, the disk
remains optically thin with respect to neutrinos, which have typical
energies of a few MeV (Fig.~\ref{fig:snapshots1}, left, upper
panel). Neutrino cooling mainly acts in
regions closely to the disk midplane, as neutrino emission tracks
density. Matter in the disk is heated as it gradually falls into the
BH potential due to angular momentum transport via MHD turbulence mediated
by the MRI. We check that the MRI is well resolved by monitoring the
wavelength of the fastest-growing MRI mode, $\lambda_\text{MRI}$, which is
typically resolved by 10 or more grid points
(cf.~Fig.~\ref{fig:snapshots1}, left); $\lambda_\mathrm{MRI}$ is
estimated by $\lambda_\mathrm{MRI}=(2\pi/\Omega) (b / \sqrt{4\pi\rho h
  + b^2})$ \cite{Kiuchi2015a},
where $\Omega$ is the angular frequency, $\rho$ the rest-mass density,
$h$ the specific enthalpy, and $b=\sqrt{b^\mu b_\mu}$ the comoving
magnetic field strength. Very close to the BH resolving the MRI
becomes challenging with current computational resources and
$\lambda_\mathrm{MRI}$ is not resolved by $>10$ grid points at all
times and spatial points. At the beginning of the simulation,
after initial amplification by
magnetic winding, the onset of the MRI further amplifies the weak
initial seed magnetic field in the disk over a few rotational periods
(resulting in a total amplification of roughly two orders of magnitude
for the maximum field strength), before the disk settles into a
saturated MRI state. Triggering the MRI both in the poloidal and
toroidal components entirely without magnetic winding (for the
same initial seed field strength) would require higher resolution and
would thus be challenging with current computational resources; this
simulation only represents a first attempt in this direction.
We note that the resulting typical magnetic field
strengths of up to $\sim\!10^{15}$ G close to the BH and the midplane,
and typical magnetic-to-fluid pressure ratios of
$\sim\!10^{-3}-10^{-1}$, are
similar to values found in early
BNS post-merger accretion systems
\cite{Ciolfi2017a,Kiuchi2015a}. Figure \ref{fig:energy_ratio} shows
the evolution of the (density-averaged) ratio of
electromagnetic energy to internal energy of the disk,
$\langle e_\mathrm{EM}/e_\mathrm{int}\rangle_D = \langle n_\mu n_\nu
T^{\mu\nu}_\mathrm{EM}/\epsilon\rho W\rangle_D$, where $n^\mu$
denotes the unit normal to the spatial hypersurfaces of the
spacetime foliation, $T^{\mu\nu}_\mathrm{EM}$ the stress-energy
tensor of the electromagnetic field, $\epsilon$ the
specific internal energy, $W$ the Lorentz factor, and $\langle
\cdot\rangle_D\equiv \int \cdot D \mathrm{d}^3x / \int
D \mathrm{d}^3x$, with $D=\sqrt{\gamma}\rho W$ the conserved
rest-mass density and $\gamma$ the determinant of
the spatial metric. This ratio stays roughly constant for
$t>20\,\mathrm{ms}$ (in a time-averaged sense) and thus indicates that a
steady turbulent state of the disk is indeed achieved.

Optically thin neutrino cooling in the midplane is balanced by
MHD-driven heating, and the disk regulates itself to a mildly
degenerate state with low $Y_\text{e}$ \cite{Beloborodov03}. The
latter results from a negative feedback process: higher electron degeneracy
$\mu_\text{e}/k_\text{B}T$ results in less electrons (lower $Y_\text{e}$)
and positrons, causing less neutrino emission, i.e., a lower cooling
rate, therefore higher temperatures, and thus lower
degeneracy; the resulting state is independent of the initial
conditions. Figure~\ref{fig:snapshots2} shows the disk once it has fully
self-regulated itself into this mildly degenerate state
($\mu_\text{e}/k_\text{B}T \sim 1$). The inner disk
remains neutron rich ($Y_\text{e}\approx 0.1$) over the course of the
simulation up to radii $r\lesssim\!60\,\text{km}$ ($\lesssim\!14$
gravitational radii), consistent with previous one-dimensional models
of neutrino-cooled disks \cite{Chen2007,Metzger+09a}.

Above the disk midplane powerful thermal outflows are generated. 
These are the result of a heating-cooling imbalance: in regions
of lower density, viscous heating from
MHD driven turbulence and energy release from recombination of free
nucleons into alpha particles exceeds cooling by neutrino emission,
and the weak interactions essentially `freeze-out' (although further
mixing can still change $Y_\text{e}$). In the polar funnel these
outflows possess high-$Y_\text{e}$
($>0.2$) and high specific-entropy
($s\gtrsim\!100\,k_\text{B}/\text{b}$), while the denser
equatorially-directed outflows have lower specific entropy
($\sim\!10\,k_\text{B}/\text{b}$) and lower $Y_\text{e}$.

Thermodynamic properties of the outflow are recorded by $10^4$ passive
tracer particles that are advected with
the fluid. We place these tracer particles of equal mass in the
initial setup with a probability proportional to the conserved
rest-mass density $D=\sqrt{\gamma}\rho W$.
Tab.~\ref{tab:results} and Fig.~\ref{fig:histograms} characterize the
outflow properties relevant to the r-process, including $Y_\text{e}$,
$s$, and the expansion timescale
$t_\text{exp}=r/v$, where $v$ denotes the
three-velocity (e.g., \cite{Lippuner2015}). These
quantities are evaluated for each
tracer particle at the last
time $t=t_{5\text{GK}}$ when the temperature of the particle drops below
5\,GK. At 5\,GK, NSE breaks
down and full nuclear reaction network calculations are required to
track nuclear abundances. We distinguish between total outflow,
defined as all tracer particles that have reached $r\ge
10^{3}\,\text{km}$ by the end of the simulation, and unbound outflow,
defined as those that are additionally unbound
according to the Bernoulli criterion $-h u_t > 1$, where $u_t$ is the time-component of the four-velocity.

By the end of the simulation, $\approx\! (16-23)\%$ of the initial
disk mass has been ejected into unbound outflows with $v\approx
(0.03-0.1)c$. With the disk still launching outflows by the end of
the simulation, our GRMHD setup potentially unbinds significantly more
mass compared to two-dimensional, non-MHD, Newtonian simulations with
similar disk parameters (Tab.~\ref{tab:results};
\cite{Fernandez2015a,Just2015a}). Polar outflows show higher
$Y_\text{e}$ and specific entropy, and smaller
$t_\text{exp}$ than equatorial outflows,
consistent with
\cite{Fernandez2015a}, while we find a factor
$\sim\!20$ higher mass in polar outflows. Our $Y_\text{e}$
distribution shows a smaller mean and does not extend as high as in
\cite{Fernandez2015a,Just2015a}.  This may indicate that neutrino
absorption (not included here) plays a dominant role in setting the
high-$Y_\text{e}$ tail of the distribution. Indeed, a preliminary
re-analysis including effects of neutrino absorption as in
\cite{Roberts2017a} shows the ejecta achieves a high-$Y_\text{e}$ tail
extending up to $\gtrsim\!0.3$; however, our finding of a sizable
quantity of low-$Y_\text{e}$ ejecta, capable of a full (2nd and 3rd peak)
r-process, remains robust. Alternatively, previously employed
pseudo-Newtonian potentials and $\alpha$-disks may not accurately
capture the heating/cooling interplay which also controls the
evolution of $Y_\text{e}$.

Full nuclear reaction network calculations with SkyNet
\cite{Lippuner2015} were performed on the tracer particles in a
post-processing step, starting at
$t=t_\text{10GK}$.  Figure~\ref{fig:nucleosynthesis} shows that the
solar abundances \cite{Arnould2007} are well reproduced
throughout the mass number ($A$) range from the 2nd r-process peak
($A\sim 130$) to the rare-earth peak ($A\sim 165$) to
the 3rd r-process peak ($A\sim 195$). There is also excellent
agreement with observed abundances in metal-poor stars
\cite{Sneden2003,Roederer2012a,Roederer2012b}. We find an
overproduction at $A=132$ as observed in
\cite{Wu2016,Lippuner2017a}. Below the 2nd r-process peak, we recover
the trends of the observed solar abundance pattern,
but overall underproduce these nuclei, which is consistent with the
absence of a significant high-$Y_\text{e}$ tail $Y_\text{e}>\!0.25$
(Fig.~\ref{fig:histograms}); however, a preliminary re-analysis
including effects of neutrino absorption as in \cite{Roberts2017a}
indicates that the entire range of r-process nuclides can be obtained.


\textit{Conclusion.---}We have shown that neutrino-cooled accretion
disks in 3D GRMHD quickly self-regulate themselves into a state of
moderate electron degeneracy (low $Y_\text{e}$) where heating from
MRI-driven turbulence is balanced by neutrino cooling. The outflows launched
self-consistently as a result of this state tend to unbind more mass
with a lower average $Y_\text{e}$ than previous axisymmetric Newtonian
simulations employing an $\alpha$-viscosity. The nucleosynthesis
yields show that these outflows are sufficiently neutron rich to
trigger a strong r-process and are well able to reproduce observed
solar abundances and observed r-process abundances in metal poor stars
from the 2nd to the 3rd r-process peak. Significant contributions to
abundances below the 2nd r-process peak can also
come from BNS mergers leading to an accretion disk
around a metastable hot neutron star, which, due to its
strong neutrino emission, may raise $Y_\text{e}$ in part of the
outflow material \cite{Metzger2014c,Wu2016,Lippuner2017a}.

\emph{Note added.}---Following the submission of this paper, a BNS merger was detected by Advanced LIGO and Virgo \cite{LIGO+17DISCOVERY}. The properties of the infrared kilonova emission observed from this event (inferred total ejecta mass $\approx 0.05M_{\odot}$ and mean velocity $v \approx 0.1\,c$; e.g. \cite{Cowperthwaite+17}) are consistent with the lanthanide-rich matter predicted in this work from disk outflows from a torus of initial mass $\approx 0.1M_{\odot}$.


\acknowledgments

We thank A.~Beloborodov, R.~Fern\'andez, R.~Haas, W.~Kastaun,
J.~Lippuner, P.~Moesta, C.~Ott, and D.~Radice for valuable discussions
throughout the course of this work. Resources supporting this work
were provided by the NASA High-End Computing (HEC) Program through the
NASA Advanced Supercomputing (NAS) Division at Ames Research
Center; the work presented here consumed a total allocation worth
$\approx\!5.7\,\text{MCPUh}$. Support for this work was provided by
the National Aeronautics and Space Administration through Einstein
Postdoctoral Fellowship Award Number PF6-170159 issued by the Chandra
X-ray Observatory Center, which is operated by the Smithsonian
Astrophysical Observatory for and on behalf of the National
Aeronautics Space Administration under contract NAS8-03060.  BDM and
DMS acknowledge support from NASA ATP grant NNX16AB30G and NSF grant AST-1410950.
 

\bibliographystyle{apsrev4-1}
\bibliography{references}

\end{document}